\documentclass[aps,pre,twocolumn,superscriptaddress,showpacs]{revtex4}
\usepackage{tabularx,graphicx}
\begin{document}
\title{Low temperature properties of a quantum particle 
coupled to dissipative environments}

\author{F. Guinea}
\affiliation{Instituto de Ciencia de Materiales de Madrid,
CSIC, Cantoblanco, E-28049 Madrid, Spain.}

\begin{abstract}
We study the dynamics of a quantum particle coupled to dissipative (ohmic)
environments, such as an electron liquid.
For certain choices of couplings, the properties
of the particle can be described in terms of an effective mass.
A particular case is the three dimensional
dirty electron liquid. In other environments, like
the one described by the Caldeira-Leggett model, the effective
mass diverges at low temperatures, and quantum effects are
strongly suppressed. For interactions within this class, arbitrarily
weak potentials lead to localized solutions. Particles
bound to external potentials, or moving in closed orbits, can show a
first order transition, between strongly and weakly localized
regimes.
\end{abstract}
\date{\today}
\maketitle
\section{Introduction}
The suppression of quantum effects by dissipative environments
is a subject of current debate. It is known that in certain cases, 
like a quantum particle tunneling in a two level system\cite{C82} or
in a periodic potential\cite{S83}, whose coupling to the environment
can be described by the Caldeira-Leggett model\cite{CL81}, quantum effects
can be almost completely suppressed at
low temperatures. On the other hand, 
perturbative calculations for the dephasing rate of electrons
in a dirty Fermi liquid suggest that dephasing becomes irrelevant
at low temperatures\cite{AAK82,SAI90}.
Recent experiments\cite{MJW97}, see also\cite{Aetal01}, have
increased the interest in this topic, and a variety of
theoretical analyses, with different conclusions, have 
followed\cite{GZ98,GZS01,AAG99,AAV01,B01,I02,GHZ02}.

The simplest situation where the effect of the dissipation
can be studied is that of a single particle 
interacting with an external environment. A particular case is 
the Caldeira-Leggett model, where a specific choice of
environment and coupling is made. This model has been extensively
studied, and it is known that the diffusion of the particle
at long times cannot be expressed in terms of a finite 
effective mass. In the following, we analyze related models
which describe environments such as a dirty electron liquid.
The results can be useful in understanding the quantum properties
of external particles, such as protons or muons, at surfaces or
inside metals\cite{G90,SP98,LH00}. In addition, the models
studied here provide the simplest examples where the effects
of the environment on quantum effects
can be studied in the strong coupling, non 
perturbative, regime.

The next section describes the general features of the model.
The method of calculation is discussed next.
 Section IV
presents the results for a particle moving in free space.
Section V includes extensions for particles localized
around external potentials,
or moving in bound orbits. A comparison with perturbation theory
is presented in Section VI.
The main conclusions of the present work are discussed in section VI.
\section{The model}
A convenient scheme to treat the problem 
of a particle interacting with an external environment is to
integrate out the environment, and to analyze the resulting
dynamics of the particle using the path integral formulation
of quantum mechanics. The effect of the environment is to induce
a retarded interaction among different positions
${\bf \vec{X}} ( \tau ) ,  {\bf \vec{X}} ( \tau' )$ along 
a given path. 
The coupling to a given environment is defined as ohmic if
this interaction decays as $(\tau - \tau')^{-2}$ at
long times.  
In this case, the dynamics of the particle in
the classical limit, $\hbar^2 / M \rightarrow 0$, where $M$
is the mass of the particle, can be described in terms of
a macroscopic friction coefficient, $\eta$, which is finite as
the velocity of the particle approaches zero (see below).

The effective interaction mediated by the environment can be
expressed in terms of the response function of the environment, assuming
that the coupling of the particle to each individual excitation
is weak\cite{CL81}, or, alternatively, that the environment is
weakly perturbed. In the following, we assume that the
environment is an electron liquid, and that the particle
is coupled, by a local potential, to the electronic charge 
fluctuations\cite{G84}:
\begin{equation}
{\cal H}_{int} =  \int d^3 {\bf \vec{x}} 
V \left(
{\bf \vec{x}} - {\bf \vec{X}} \right) \rho ( {\bf \vec{x}} )
\label{coupling}
\end{equation}
where ${\bf \vec{X}}$ is the coordinate of the particle,
$\rho ( {\bf \vec{x}} )$ describes the charge fluctuations
of the electrons, and the $V$ is the coupling potential.
The induced retarded putential can be written as:
\begin{widetext}
\begin{equation}
{\cal V} \left[ {\bf \vec{X}} ( \tau ) - {\bf \vec{X}}
( \tau' ) \right] = \int d {\bf \vec{q}} \int d \omega
e^{i {\bf \vec{q}} \left[ {\bf \vec{X}} ( \tau ) - {\bf \vec{X}}
( \tau' ) \right]} e^{i \omega ( \tau - \tau' )} V^2 
( {\bf \vec{q}} ) {\rm Im} \chi ( {\bf \vec{q}} ,
\omega )
\label{retarded}
\end{equation}
\end{widetext}
where ${\rm Im} \chi ( {\bf \vec{q}} ,
\omega )$ is the Fourier transform of:
\begin{equation}
{\rm Im} \chi ( {\bf \vec{x}} - {\bf \vec{x}'} , \tau - \tau' )
= \langle \rho ( {\bf \vec{x}} , \tau ) 
\rho ( {\bf \vec{x}'} , \tau' ) \rangle
\end{equation}
For an electron liquid
we have:  $\lim_{\omega \rightarrow 0}
{\rm Im} \chi ( {\bf \vec{q}} ,
\omega ) \propto | \omega |$, and that fixes the long time
behavior of the retarded interaction, eq.(\ref{retarded}),
which decays as $( \tau - \tau' )^{-2}$. Finally, we can write
for the effective action of the particle\cite{G84,SG87}:
\begin{widetext}
\begin{equation}
S = \int d \tau \frac{M}{2}
\left( \frac{\partial {\bf \vec{X}}}{\partial \tau}
\right)^2 + \int d \tau d \tau'
\frac{{\cal F} \left[ \left| {\bf \vec{X}} ( \tau ) - {\bf \vec{X}} ( \tau' )
 \right|^2 \right]}{| \tau - \tau' |^2}
\label{action}
\end{equation}
\end{widetext}
The function ${\cal F}$ is determined by the coupling
between the particle and the electronic charge fluctuations.

The Caldeira-Leggett model can also be written in the
form given in eq.(\ref{action}), where
the function ${\cal F}$ 
is ${\cal F} \propto \eta  | {\bf \vec{X}} ( \tau ) -
{\bf \vec{X}}' ( \tau' ) |^2$, and
$\eta$ is the macroscopic friction coefficient.
Other generalizations of the Caldeira-Leggett model, 
used in relation to Coulomb blockade, include higher
order terms in the collective coordinate\cite{GZ01,Fetal02},
associated with higher order tunneling processes. These terms
typically involve many different times as well. The derivation
of eq.(\ref{action}), using second order perturbation
theory, leads to non linear terms which couple the coordinates
at two different times only.

In general, we can write the function ${\cal F}$ as:
\begin{equation}
{\cal F} ( u )  = \alpha  f \left( \frac{u}{l^2} \right)
\label{spatial}
\end{equation}
where $\alpha$ is a dimensionless constant and $l$ is a length
scale typical of the fluctuations in the environment.
We assume, without loss of generality,
that $f ( 0 ) = 0$ and $f' ( 0 ) = 1$.
At high temperatures, where $l_T = \sqrt{\hbar^2 / ( M T )}
\ll l$, the motion of the particle is determined 
by $\lim_{u \rightarrow 0} {\cal F} ( u ) \approx
\alpha / l^2 f' ( 0 )$. In this limit, eq.(\ref{action})
is equivalent to the Caldeira-Leggett model with a
friction coefficient $\eta = \alpha / l^2$.

In the following, we
will assume that the response function of
the environment is that of
a dirty (diffusive) electron liquid, and the
external particle couples to the charge 
fluctuations via a screened Coulomb potential. Then:
\begin{eqnarray}
\chi_0 ( {\bf \vec{q}} , \omega ) &\approx &\nu
\frac{{\cal D} | {\bf \vec{q}} |^2}{i \omega + {\cal D} | {\bf \vec{q}} |^2}
\nonumber \\
{\cal F} ( {\bf \vec{q}} ) &\approx &\frac{1}
{\nu {\cal D} | {\bf \vec{q}} |^2}
\label{dirty}
\end{eqnarray}
where ${\cal F} ( {\bf \vec{q}} )$ is the Fourier transform
of ${\cal F} \left[ {\bf \vec{X}} ( \tau ) - {\bf \vec{X}}' ( \tau' ) \right]$.
In the calculation of ${\cal F}$ we have included the full selfconsistent
screened Coulomb potential. The expressions in eqs.(\ref{dirty})
determine the 
the function ${\cal F}$ in eq.(\ref{action}) for distances
greater than the mean free path, $l$. At shorter distances,
we will choose a regularization consistent with the
expected asymptotic regime. Finally, we will compare the
results with two other choices of the retarded interaction: 
i) the one appropiate for the Caldeira-Leggett model, and
ii) a retarded interaction which decays exponentially beyond
a certain length, $l$\cite{C97}. The retarded actions to be considered,
expressed in terms of the function $f$ given in eq.(\ref{spatial}),
are: 
\begin{eqnarray}
& & \nonumber \\
f ( u ) &= u &{\rm  \, \, Caldeira-Leggett \, \,  model} \nonumber \\
f ( u ) &= 1 - e^{-u} &{\rm  \, \, short \, \, range \, \, kernel} \nonumber \\
f ( u ) &= 2  \sqrt{1 + u} - 2  &{\rm  \, \, 1D \, \, dirty \, \,
electron \, \, gas} \nonumber \\
f ( u ) &= \log ( 1 + u ) &{\rm  \, \, 2D \, \, dirty \, \,
electron \, \, gas} \nonumber \\
f ( u ) &= 2 - \frac{2}{\sqrt{1+u}} 
&{\rm  \, \, 3D \, \, dirty \, \, electron
 \, \, gas} 
\label{actions}
\end{eqnarray}
Combining eq.(\ref{spatial}) and eqs.(\ref{actions}), the macroscopic
friction coefficient is, in all cases, $\eta = \alpha / l^2$. 
For the dirty electron gas, $\alpha \approx l^{2-D} / ( 
\nu {\cal D} )$.
This value can change if the particle is coupled to different
electronic reservoirs. For a clean electron gas, $\alpha$
can be written in terms  of the phaseshifts induced by the presence
of the particle on the electrons at the Fermi level\cite{GS85}.
In the following, we will treat $\alpha$ as a variable which can take
arbitrary values.

\section{The calculation}
\subsection{Perturbative expansion.}
The action described in eq.({\ref{action}) can be solved exactly
when the retarded interaction is given by the Caldeira-Leggett
model. In this case, the action depends quadratically
on the spatial coordinate, ${\bf \vec{X}} ( \tau )$.
In the following, we will assume that $\tau$ is the imaginary
time.

In general, one can write the expansion:
\begin{widetext}
\begin{equation}
S = \int d \tau \frac{M}{2}
\left( \frac{\partial {\bf \vec{X}}}{\partial \tau}
\right)^2 + \sum_n \alpha_n \int d \tau d \tau'
\frac{\left| {\bf \vec{X}} ( \tau ) - {\bf \vec{X}} ( \tau' )
 \right|^{2n} }{| \tau - \tau' |^2}
\label{action_exp}
\end{equation}
\end{widetext}
where:
\begin{equation}
\alpha_n = \alpha \frac{1}{n! l^{2n}} \frac{\partial^n f}{\partial u^n}
\end{equation}
The case where only the $n=1$ term in the sum in eq.(\ref{action_exp})
is non zero corresponds to the Caldeira-Leggett model.
One can write a perturbative expansion for the corrections  due to the
$n > 1$ terms. Typical diagrams are given in Fig.[\ref{diagrams}].
It is easy to show that all $\alpha_n , n > 1$ acquire logarithmic
corrections. In Renormalization Group terms, they are all
marginal. In this respect, the model differs from standard models
in statistical mechanics, where usually only quartic terms need
to be considered. If we only consider the quartic term, $\alpha_2$,
the logarithmic divergence of the perturbative series lead to the
scaling equations (see Fig.[\ref{diagrams}]):
\begin{eqnarray}
\frac{\partial \alpha_1}{\partial l} &= &\frac{2 \alpha_2}{\alpha_1}
\nonumber \\
\frac{\partial \alpha_2}{\partial l} &= &- \frac{4 \alpha_2^2}
{\alpha_1^2}
\label{scaling}
\end{eqnarray}
where $l = \log ( \Lambda_0 / \Lambda )$, $\Lambda_0$ is the initial high
energy (short time) cutoff needed to regularize eq.(\ref{action}),
and $\Lambda$ is the effective cutoff. A scaling approach following
eq.(\ref{scaling}) is, however, impractical, because
one should consider an infinite set of coupled equations,
including all the couplings.
\subsection{Large N approximation.} 
The non linear terms in the action in eq.(\ref{action_exp}) are greatly
simplified when the number of components $N$ of the vector ${\bf \vec{X}}$
is large. In order to have a consistent theory, we
must rescale the argument $u$ of the function
 $f ( u )$ in eq.(\ref{spatial})  so that
$f ( u ) = N f ( \bar{u} / N )$, where $\bar{u}$ is proportional
to $| {\bf \vec{X}} ( \tau ) - {\bf \vec{X}}' ( \tau' )
|^2$. 
In terms of the diagrammatic expansion sketched in
Fig.[\ref{diagrams}], we need only to consider closed loop
diagrams, like those in Fig.[\ref{bubbles}]. 
These diagrams can be summed using standard large-N techniques
in statistical mechanics\cite{FMN72,S73,C85}, leading to the equations:
\begin{eqnarray}
\Sigma ( \tau ) &= &\frac{\alpha}{\tau^2} f' [ G ( \tau ) ] \nonumber \\
G ( \omega ) &= &\frac{l^2}{\frac{M l^2 \omega^2}{2}  + \Sigma ( \omega )}
\label{sigma}
\end{eqnarray}
where:
\begin{equation}
G ( \tau ) = \left\langle \left| {\bf \vec{X}} ( \tau ) - 
{\bf \vec{X}} ( 0 ) \right|^2 \right\rangle
\label{green}
\end{equation}
and, using eq.(\ref{sigma}), we can write:
\begin{equation}
G ( \tau ) = l^2 \bar{G} \left[ \alpha , 
\frac{\hbar^2}{2 M l^2}  \tau  \right]
\label{dimension}
\end{equation}
\begin{figure}
\includegraphics[width=6cm]{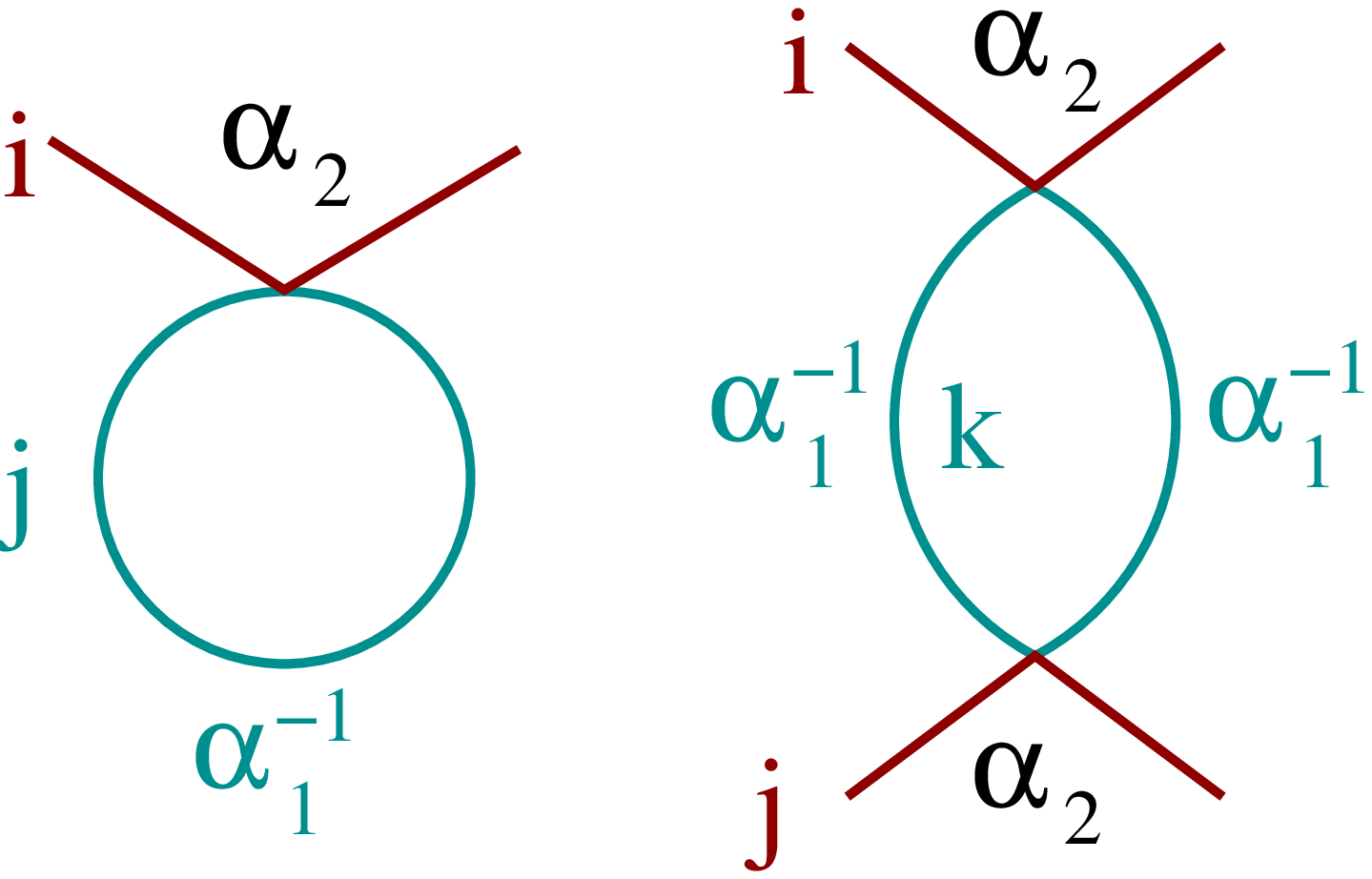}
\caption{Some simple diagrams which renormalize the values
of $\alpha_1$ and $\alpha_2$ in eq.(\protect{\ref{action_exp}}).
The propagators are proportional to $\alpha_1^{-1}$. Indices denote
components of the vector ${\bf \vec{X}}$.}
\label{diagrams}
\end{figure}
Two exactly solvable cases are:
\begin{equation}
G ( \tau ) \propto \left\{ \begin{array}{lr}
\frac{\tau }{M} &{\rm free \, \, particle \, \, of
\, \,  mass} \, \,  M \\
\frac{1}{\eta} \log \left[ \frac{\eta \tau
}{M} \right] &{\rm Caldeira-Leggett \, \,
model} \end{array}
\right.
\end{equation}

\begin{figure}
\includegraphics[width=\columnwidth]{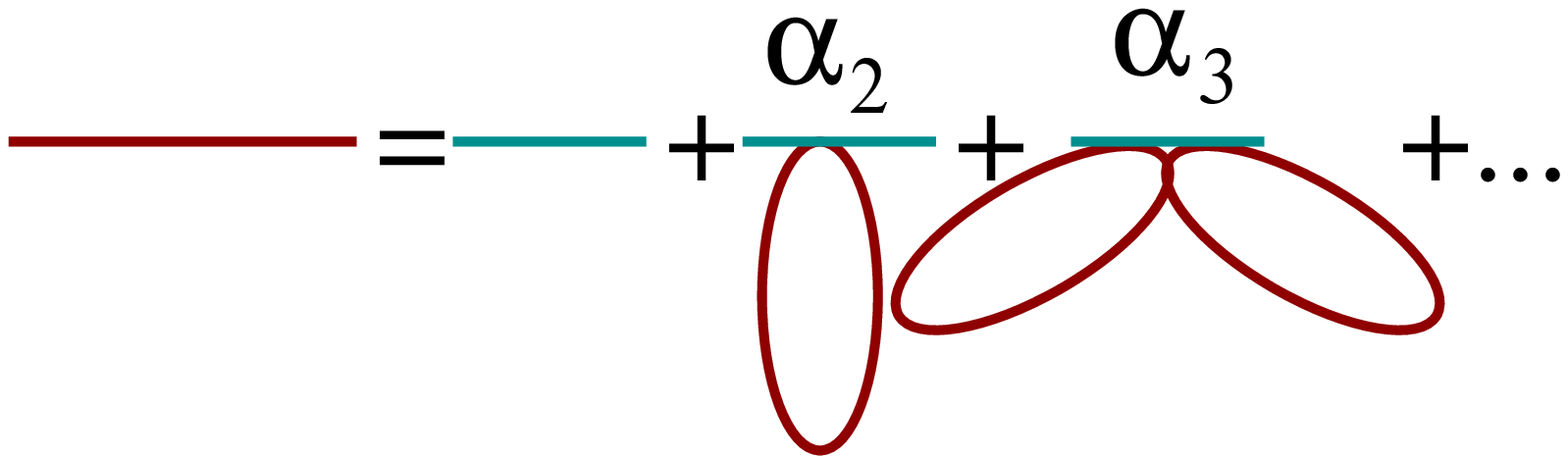}
\caption{Sketch of the selfconsistent solution of
closed loop diagrams which leads to eqs.(\protect{\ref{sigma}}).}
\label{bubbles}
\end{figure}
\subsection{Variational ansatz.}
The large-$N$ expansion described in the previous subsection
can be alternatively formulated as a variational approximation
to the action in eq.(\ref{action}). We use the ansatz:
\begin{widetext}
\begin{equation}
S_0 = \int d \tau \frac{M}{2}
\left( \frac{\partial {\bf \vec{X}}}{\partial \tau}
\right)^2 + \int d \tau d \tau'
\left| {\bf \vec{X}} ( \tau ) - {\bf \vec{X}} ( \tau' )
 \right|^2 \Sigma ( \tau - \tau')
\label{action_var}
\end{equation}
\end{widetext}
where $\Sigma ( \tau )$ is a function to be determined from
the minimization of $\langle S - S_0 \rangle_0 + F_0$.
The subscript 0 means averaging with respect to $S_0$,
and $F_0$ is the free energy associated to $S_0$.

The function $\Sigma ( \tau )$ satisfies the equation:
\begin{equation}
\Sigma ( \tau ) = \frac{1}{\tau^2}
\frac{\partial}{\partial G_0 ( \tau  )} 
\left\langle 
{\cal F} \left[ \left| {\bf \vec{X}} ( \tau ) - {\bf \vec{X}} ( 0 )
\right|^2 \right] \right\rangle_0
\label{Sigma}
\end{equation}
and:
\begin{equation}
G_0 ( \tau ) = \left\langle
\left| {\bf \vec{X}} ( \tau ) - {\bf \vec{X}} ( 0 )
\right|^2 \right\rangle_0
\label{g0}
\end{equation}
Eqs. (\ref{Sigma}) and (\ref{g0}) are equivalent to eqs.(\ref{sigma}).
The advantage of the variational formulation is that it can be extended
to finite values of $N$. The expectation values to be calculated
are of the type:
\begin{widetext}
\begin{equation}
\left\langle
{\cal F} \left[ \left| {\bf \vec{X}} ( \tau ) - {\bf \vec{X}} ( 0 )
\right|^2 \right] \right\rangle_0 = C_N
\int_0^{\infty} d r r^{N-1} {\cal F} ( r^2 )
e^{- r^2 / [ 2 G_0 ( \tau ) ]}
\label{integral_2}
\end{equation}
\end{widetext}
where:
\begin{eqnarray}
C_N = &\sqrt{\pi} \sqrt{G_0 ( \tau - \tau' )}  &N = 1 \nonumber \\
C_N = &\sqrt{\pi} [ G_0 ( \tau - \tau' ) ]^{3/2}  &N = 3 \nonumber \\
\end{eqnarray}
\begin{figure}
\includegraphics[width=6cm]{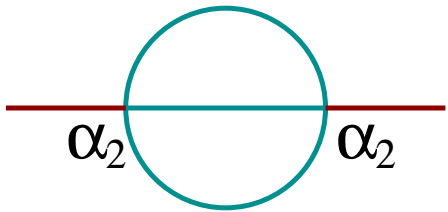}
\caption{Simplest diagram
which gives a vertex correction to the variational scheme discussed
in the text.}
\label{vertex}
\end{figure}
\subsection{Vertex corrections.}
For the physically relevant class of $N=2,3$, the above approximations
neglect vertex corrections, like those shown in Fig.[\ref{vertex}].
In the next section, we will analyze them, and argue that they do not
change qualitetively the solutions.
\section{Results. Free particle}
\subsection{Parameters of the model.}
Eqs. (\ref{sigma}) can be solved selfconsistently. The model
has two dimensionless parameters, the value of $\alpha$,
defined in eqs.(\ref{spatial},\ref{actions}), and the ratio
$\hbar^2 / ( M l^2 \Lambda_0 )$, where $\Lambda_0$ is the
cutoff in the spectrum of the environment, which we will
assume to be finite.
Note, however, that, for most physical quantities, the limit
$\hbar^2 / ( M l^2 \Lambda_0 ) \rightarrow 0$ is well defined.
The scaling given in eq.(\ref{dimension}) allows us to
set $M = 1$ and $l = 1$.
\subsection{Asymptotic analysis.}
We first analyze the selfenergy correction due to
the environment, $\Sigma ( \omega )$ in eq.(\ref{sigma}).
At very short times, $\tau \ll \tau_0 \Lambda_0^{-1}$, the environment
cannot influence the particle, and $\Sigma ( \tau ) = 0$.
For times $\tau \gg \tau_0$ such that $G ( \tau )$ is small,
$\left\langle
\left| {\bf \vec{X}} ( \tau ) - {\bf \vec{X}} ( \tau' )
\right|^2 \right\rangle \ll l^2$. Then: 
\begin{equation} 
{\cal F} \left( \left| {\bf \vec{X}} \right|^2 \right)
\sim \alpha \frac{\left| {\bf \vec{X}} \right|^2}{l^2}
\end{equation}
In this regime, the behavior of the system is indistinguishable
from that of an effective Caldeira-Leggett model, and:
\begin{eqnarray}
\Sigma ( \omega ) &\approx &\frac{\alpha}{l^2} | \omega | \nonumber \\
G ( \tau ) &\approx &\frac{l^2}{\alpha} \log \left(
\frac{\alpha \tau}{M l^2} \right) 
\label{short_time}
\end{eqnarray}
This approximation is valid provided that $G ( \tau ) \ll l^2$.
This contraint sets a maximum time, $\tau_1 \sim ( M l^2 e^{\alpha}) / 
\alpha$, beyond which the diffusion of the particle
cannot be described by eqs.(\ref{short_time}).
For $\tau \gg \tau_1$, $\Sigma ( \omega )$ acquires a contribution:
\begin{equation}
\Sigma ( \omega ) = \int_{\tau_0}^{\tau_1} \frac{\alpha}{l^2}
\frac{1 - e^{i \omega \tau}}{\tau^2} + \Sigma_{\tau \gg \tau_1} (
\omega ) \sim
\frac{\alpha}{l^2} \omega^2 \tau_1 + \Sigma_{\tau \gg \tau_1}
\label{sigma_short} ( \omega )
\end{equation}
where a a term arising from times longer than $\tau_1$,
$\Sigma_{\tau \gg \tau_1} ( \omega )$, has
to be added.

$\Sigma_{\tau \gg \tau_1} ( \omega )$ is determined by
$\lim_{u \rightarrow \infty} f ( u )$ in eqs.(\ref{actions}).
If $\Sigma_{\tau \gg \tau_1} ( \omega )$ 
goes to zero faster than $\omega^2$ as $\omega \rightarrow 0$,
then $G ( \tau ) \propto \tau$.
Hence, we can check if the asumption that $G ( \tau ) \propto \tau$
leads to a self consistent solution. We insert this ansatz
into the expression of $\Sigma_{\tau \gg \tau_1}
 ( \tau )$. At long times, we have:
\begin{equation}
\lim_{\tau \rightarrow \infty} \Sigma ( \tau )
\propto \frac{\alpha}{\tau^2} \lim_{\tau \rightarrow \infty} f'
[ G ( \tau ) ] 
\end{equation}
where $c$ is a constant. If $f' [ G ( \tau ) ]$ decays faster than
$G ( \tau )^{-1}$, then $\lim_{\omega \rightarrow 0}
\Sigma_{\tau \gg \tau_1} ( \omega )$ goes to zero faster
than $\omega^2$. 

Thus, if $\lim_{u \rightarrow \infty} f' ( u )$
decays faster than $u^{-1}$, we can write:
\begin{widetext}
\begin{equation}
G ( \tau ) \sim \int_{0}^{\tau_1^{-1}}
d \omega \frac{1 - e^{i \omega \tau}}{M \omega^2 + \omega^2 \tau_1 / l^2}
\sim l^2 \frac{\tau}{\tau_1} \sim M e^\alpha \tau
\end{equation}
\end{widetext}
This equation describes the propagation of a quantum particle
with effective mass $M_{eff} \sim M e^\alpha$.

In terms of the action in eq.(\ref{action}),
the previous analysis allows us to describe the dynamics of the
quantum particle in terms of a renormalized effective mass when
the function which defines the coupling to the environment is
such that:
\begin{widetext} 
\begin{equation}
\lim_{\left| {\bf \vec{X}} ( \tau ) - {\bf \vec{X}} ( \tau' )
\right|^2 \rightarrow \infty}
\left\langle
{\cal F} \left[ \left| {\bf \vec{X}} ( \tau ) - {\bf \vec{X}} ( \tau' )
\right|^2 \right] \right\rangle  \sim o \left[
\left| {\bf \vec{X}} ( \tau ) - {\bf \vec{X}} ( \tau' )
\right|^{-2} \right]
\label{bound}
\end{equation}
\end{widetext}
This is the case when the environment is a three dimensional
dirty electron liquid, or for a short range kernel, 
see eqs.(\ref{actions}). It is interesting to note that, for
finite dimensions $N$, eq.(\ref{integral_2}) implies that 
$ \left\langle
{\cal F} \left[ \left| {\bf \vec{X}} ( \tau ) - {\bf \vec{X}} ( \tau' )
\right|^2 \right] \right\rangle$ never decays faster than
$\left| {\bf \vec{X}} ( \tau ) - {\bf \vec{X}} ( \tau' )
\right|^{-N}$.
\subsection{Vertex corrections.}
The scheme used here neglects vertex corrections. 
Diagrams such as the one shown in Fig.[\ref{vertex}] give a contribution,
$\Sigma_{vertex} ( \tau )$ to $\Sigma ( \tau )$, and can change
the above results if they decay at long
times more slowly than $\tau^{-3}$. 
The diagram in Fig.[\ref{vertex}] goes as:
\begin{equation}
\Sigma_{vertex} ( \tau ) \sim \left\{ f''
\left[ G ( \tau ) \right] 
\right\}^2 G^3 ( \tau )
\end{equation}
If $G ( \tau ) \sim \tau$ and $f$ is such that eq.(\ref{bound}) is 
satisfied, then $\Sigma_{vertex} ( \tau )$ decays faster than
$\tau^{-3}$ at long times, and the correction from the
diagram in Fig.[\ref{vertex}] does not change the results
described above. It is easy to show that the same is also
valid for more complicated vertex diagrams. This analysis leads
us to conjecture that, when the dynamics of the particle can
be described in terms of an effective mass, vertex corrections
are irrelevant.
\subsection{Simple actions.}
The previous analysis allows us to study the simple case where
${\cal F} ( {\bf \vec{X}} ) = \alpha_n | {\bf \vec{X}} |^{2n}$.
The case $n=1$ corresponds to the Caldeira-Leggett model. 
For $n > 1$, we can make the ansatz $\lim_{\tau 
\rightarrow \infty} G ( \tau ) \propto a_n
\log^{\gamma_n} ( \tau )$.  Then, from eqs.(\ref{sigma}):
\begin{equation}
\Sigma ( \tau ) \propto \frac{\alpha_n}
{\tau^2} \left[ a_n \log^{\gamma_n} ( \tau ) \right]^{n-1}
\end{equation}
and:
\begin{equation}
G ( \omega ) \propto \frac{1}{\alpha_n | \omega | \left[ a_n
\log^{\gamma_n} ( \omega ) \right]^{n-1}}
\end{equation}
so that:
\begin{eqnarray}
\frac{1}{\alpha_n a_n^{n-1}} &= &a_n \nonumber \\
1 + ( 1 - n ) \gamma_n &= &\gamma_n
\end{eqnarray}
and, finally:
\begin{equation}
\lim_{\tau \rightarrow \infty}
G ( \tau ) \propto \left[ \frac{1}{\alpha_n}
\log ( \tau ) \right]^{1/n}
\end{equation}
This correlation function implies that the particle is never localized,
although it can diffuse more slowly than 
for the Caldeira-Leggett model. The effective mass
diverges at low temperatures. The case $n=1/2$ corresponds to the 
one dimensional dirty electron gas, described in eq.(\ref{actions}).
\subsection{Numerical results.}
We have solved iteratively eqs.(\ref{sigma}) for different choices
of the environment. The resulting correlation function $G ( \tau )$
is shown in
Figure[\ref{d_infty}] for the cases:
\begin{eqnarray}
f ( x ) = &\frac{1}{\sqrt{1+x}} &  
{\rm ( 3D \, \, dirty \, \, electron \, \, gas )}  \nonumber \\
f ( x ) = &\log( 1+x ) &  
{\rm ( 2D \, \, dirty \, \, electron \, \, gas )} \nonumber \\
f ( x ) = &x &  {\rm ( Caldeira-Leggett \, \, model )} \nonumber \\
\label{comp}
\end{eqnarray}
The dimensionless
parameters in the three cases are $\alpha=2$ and
$\hbar^2 / ( M l^2 \Lambda_0 )
= 0.05$. The units are such that $M = 1$ and $l = 1$.
\begin{figure}
\resizebox{8cm}{!}{\rotatebox{0}{\includegraphics{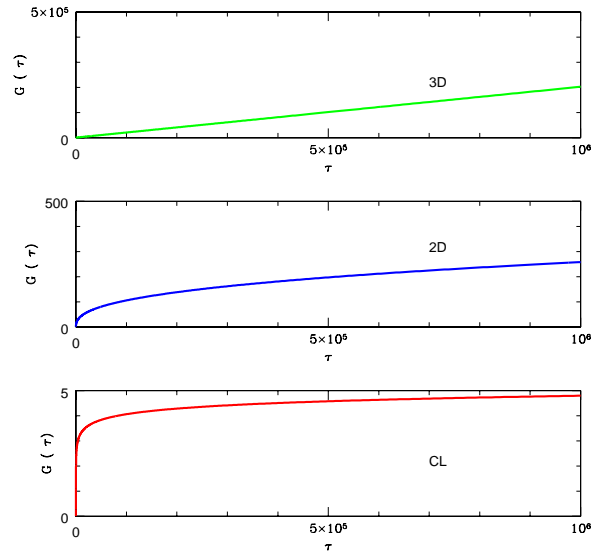}}}
\caption{Time dependence of the propagator $G ( \tau )$, for 
$\hbar^2 / ( M l^2 ) = 1$, $l = 1$ and $\alpha = 2$ for the three
cases described in eq.(\protect{\ref{comp}}).}
\label{d_infty}
\end{figure}
\begin{figure}
\resizebox{8cm}{!}{\rotatebox{0}{\includegraphics{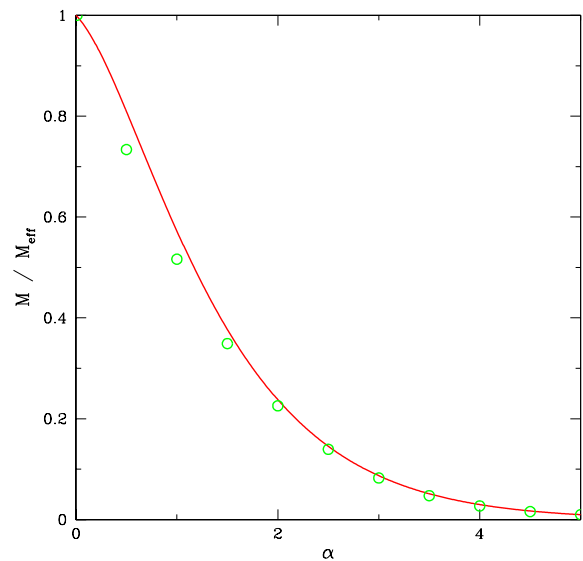}}}
\caption{Dots: Effective mass of a particle interacting with a dirty
three dimensional electron liquid, as function of the strength
of the interaction, $\alpha$. The full line is
a fit to a function of the type $M / M_{eff} = ( c_1 + c_2 \alpha )
e^{ - c_3 \alpha}$.}
\label{mass}
\end{figure}
\begin{figure}
\resizebox{8cm}{!}{\rotatebox{0}{\includegraphics{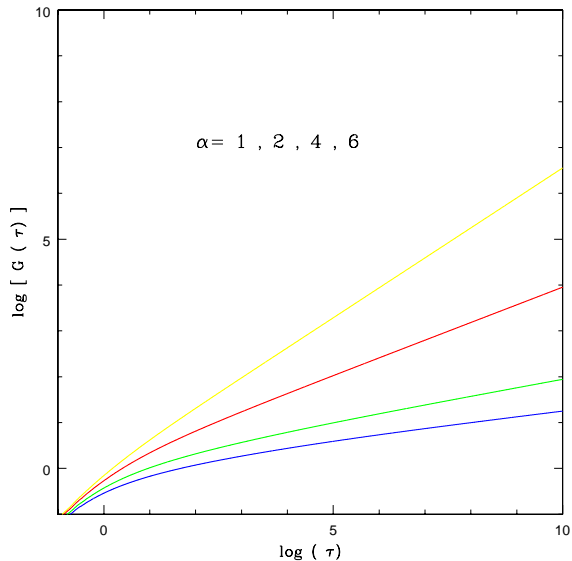}}}
\caption{Log-log plots of $G ( \tau )$ vs. $\tau$ for an action which
describes a two dimensional dirty electron liquid, 
eq.(\protect{\ref{actions}}). From top to bottom: $\alpha =
1 , 2 , 4 , 6$.}
\label{2D}
\end{figure}
The macroscopic friction coefficient, $\eta = \alpha / l^2$ is
the same in the three cases.
It is clear that the spatial dependence of the retarded interaction
induces significant differences in the long time dynamics of
the particle, although look the same at sufficiently high 
temperatures, where the mass, and the macroscopic friction coefficient
are the only relevant parameters.

For the Caldeira-Leggett model, $G ( \tau )$ increases logarithmically 
at long times. The effects of the three dimensional dirty electron gas
can be described in terms of an effective mass, which, for the
parameters used here, is about two orders of magnitude larger than the
bare mass. The two dimensional dirty electron gas is a marginal
case, and the numerical results, like those shown
in Fig.[\ref{2D}],  suggest that $G ( \tau )
\propto \tau^\kappa$, with $0 < \kappa < 1$, and $\kappa \propto
1 - g^{-1}$, where $g$ is the conductance. 

The effective mass, when the environment is described by the
dirty three dimensional liquid, as function of the strength
coupling, $\alpha$ is shown in Fig.[\ref{mass}]. The numerical results
support the exponential dependence on $\alpha$ discussed in the
preceding subsection.
\section{Results. Localized solutions}
\subsection{Local attractive potential.}
An external potential, $V$, acting on the particle can be described,
using the path integral formulation, by adding a term
to the action in eq.(\ref{action}):
\begin{equation}
S_{pot} = \int d \tau V \left[ {\bf \vec{X}} ( \tau ) \right]
\label{action_pot}
\end{equation}
This term leads to new nonlinear effects. We can extend the
large-$N$ or variational scheme described earlier by using
the approximate action:
\begin{equation}
S_{var} = \int d \tau \frac{\lambda}{2} \left| {\bf \vec{X}} ( \tau )
\right|^2
\label{lambda}
\end{equation}
where $\lambda$ is a variational parameter. This ansatz, in the
absence of dissipation, is equivalent to the use of a variational
set of gaussian wavefunctions in order to probe the existence
of bound states of the potential $V ( {\bf \vec{X}} )$. It is well
known that this method can be used to study the existence of
localized solutions for arbitrarily small potentials in one
dimension, while a threshold is required for the existence
of a bound state in three dimensions.

We now analyze the full action, given by eqs.(\ref{action}) and
(\ref{action_pot}) using the variational ansatz given by
eqs.(\ref{action_var}) and (\ref{lambda}). We will focus on the
possible existence of localized solutions, characterized
by a finite value of $\lambda$, in the limit of a very weak
potential $V ( {\bf \vec{X}} )$. Then, $\lambda l^2 \ll
\Lambda_0 , \hbar^2 / ( M l^2 )$, and the Fourier
transform of $G ( \tau )$ is:
\begin{equation}
G ( \omega ) = \frac{l^2}{\frac{M l^2 \omega^2}{2}  + 
\Sigma_{\lambda = 0} ( \omega ) + \lambda l^2}
\label{green_lambda}
\end{equation}
where $\Sigma_{\lambda = 0} ( \omega )$ is the self energy 
calculated when $\lambda = 0$. 

We now write the external potential as $V ( {\bf \vec{X}} ) = 
\bar{V} ( u )$, where $u = | {\bf \vec{X}} |^2 $.
The self consistency equation for $\lambda$ is:
\begin{equation}
\lambda = \frac{\partial}{\partial G_\infty } 
\left\langle \bar{V} \left( 
G_\infty  \right) \right\rangle_{var}
\label{lambda_eq}
\end{equation}
The expectation value in the r.h.s. of this equation is to be
calculated using the variational action, eqs.
(\ref{action_var}) and (\ref{lambda}). This calculation is
similar to that performed in eq.(\ref{integral_2}).
We are interested in the limit when $G_\infty  \gg l^2$.
Then, for localized potentials:
\begin{equation}
\frac{\partial}{\partial G_\infty }
\left\langle \bar{V} \left(
G_\infty  \right) \right\rangle_{var} \sim
\frac{1}{G_\infty^{N/2+1}}
\label{g_inf}
\end{equation}
For the Caldeira-Leggett model, $G_\infty \propto 
\eta^{-1}  \log [ \eta^2 / ( M \lambda  ) ]$. 
Combining eqs.(\ref{lambda_eq}) and (\ref{g_inf}), we find that
there is a localized solution with a
for any attractive
potential $V ( {\bf \vec{X}} )$. The corresponding value of
$\lambda$ is:
\begin{equation}
\lambda \sim \left( \frac{\eta}{\hbar} \right)^{N/2+1} 
\int d^N {\bf \vec{X}} 
V ( {\bf \vec{X}} )
\end{equation}
The value $\Delta = \lambda / \eta$ can be interpreted as a
gap in the spectrum of the excitations of the particle.

If the dynamics of the particle
can be described in terms of an effective mass, $G_\infty \propto
\sqrt{\hbar^2 / ( M_{eff} \lambda )}$.
Then, eqs.(\ref{lambda_eq}) and (\ref{g_inf}), lead to:
\begin{equation}
\lambda \propto \lambda^{N/4 + 1/2} \int d^N {\bf \vec{X}}
V ( {\bf \vec{X}} )
\end{equation}
This equation has solutions for arbitrarily weak potentials
only if $N=1$, that is, if the particle moves in one dimension, 
in agreement with standard quantum mechanics.

\subsection{Closed orbits.}
We can use the previous method to analyze the motion of the particle
when it moves around closed circular loops. When the loop is threaded
by a magnetic flux, this geometry can be used
to analyze quantum interference effects\cite{G02,GHZ02}. The trajectory of
the particle can be described in terms of an angle, and $| 
{\bf \vec{X}} ( \tau ) | = R$, where $R$ is the radius
of the orbit. The action in eq.(\ref{action}) describes
a one dimensional non linear sigma model with long
range interactions. The case when only quadratic
terms are kept in the expansion in
eq.(\ref{action_exp}) has been extensively discussed
in the literature\cite{K77,GS85}. The large-$N$ extension of
this model can be analyzed using standard techniques\cite{R97,AD02}.
The constraint can be incorporated by means of a Lagrange multiplier,
$\lambda ( \tau )$, whose fluctuations can be neglected
in the large-$N$ limit. This term in the action is given
by eq.(\ref{lambda}), and we can apply a similar scheme to that
discussed in the previous subsection. The value of $\lambda R^2$
gives the new energy scale which describes the dynamics
of the particle around a closed loop. Alternatively,
$\lambda R^2$ can be
interpreted as an inverse correlation time\cite{R97}.
This approach describes well the phase transition of the model
when the retarded interaction decays as 
$\tau^{-2 + \epsilon}$\cite{R97,AD02}.

The dynamics of the particle, using  the full action eq.(\ref{action_exp}), 
is determined by $G_{\infty}$, for $R \gg l$. The value of
$\lambda$ is given by the solution of:
\begin{equation}
G_{\infty} \approx R^2
\end{equation} 
For the Caldeira-Leggett model, this equation leads to
$\lambda \propto e^{- ( \eta R^2 ) / \hbar}$, as
$R / l \rightarrow \infty$, in agreement
with previous calculations\cite{GZ98b,G02}.
If the motion of the particle can be described in terms of an effective
mass, $\lambda R^2 \sim \hbar^2 / ( M_{eff} R^2 )$, also
in agreement with earlier work\cite{G02}.
\begin{figure}
\resizebox{8cm}{!}{\rotatebox{0}{\includegraphics{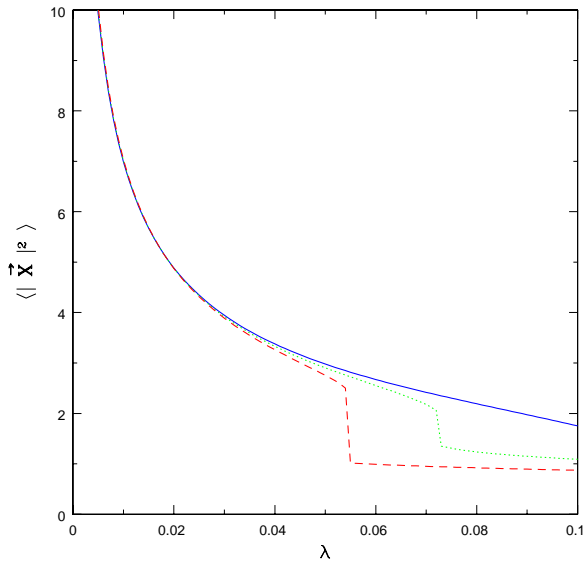}}}
\caption{First order transition obtained for $\Lambda_0 / (
\hbar^2 / M l^2 ) = 1$. The parameters are $M = 1 $ and $l = 1$,
and $f ( u ) = e^{-u}$ in eq.(\protect{\ref{spatial}}). Full curve:
$\alpha = 6$. Dotted curve, $\alpha = 8$. Broken curve:
$\alpha = 10$.}
\label{firstorder}
\end{figure}
\begin{figure}
\includegraphics[width=5cm]{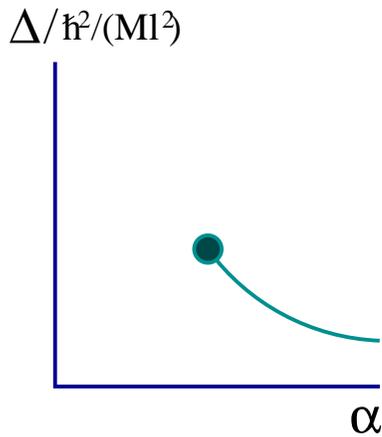}
\caption{Sketch of the transition discussed in the text.}
\label{phased}
\end{figure}
\subsection{First order phase transition.}
The results discussed in the previous subsections are valid if the
high energy cutoff of the bath, $\Lambda_0$, is much larger
than the energy scale $\hbar^2 / ( M l^2 )$. When these scales
become comparable, we find a first order phase transition,
as function of $\lambda$ or $R$ between two regimes:

i) The particle moves away from the region $| {\bf \vec{X}} | < l$
at times shorter than $\Lambda_0^{-1}$, and, for longer times, it
diffuses like a free particle.

ii) The particle is trapped inside the region $| {\bf \vec{X}} | < l$
for all times.

This transition is analogous to that described recently in the two dimensional
non linear sigma model\cite{BGH02,CP02,ES02}. The transition takes place
between two disordered phases, and it does not violate the Mermin-Wagner
theorem, which can also be formulated for the
model studied here.
The phase diagram is sketched in Fig.[\ref{phased}]. There is a
line of first order transitions, which ends at a critical point.
The discontinuity at the phase transition increases as the minimum in the
function ${\cal F} ( {\bf \vec{X}} )$ becomes deeper and more
localized around ${\bf \vec{X}} = 0$. 

The existence of a critical point, shown in Fig.[\ref{phased}],
implies the existence of physical quantities with anomalous
decay in the time domain. Following the results in\cite{CP02},
we conjecture that the energy-energy correlations will show 
power law correlations in time\cite{NB02}.
\section{Comparison with perturbation theory}
It is instructive to compare the results presented in the previous
sections with a perturbative calculation, which is the scheme most
widely used when studying dephasing\cite{AAK82,SAI90,AAG99}. 
We consider a particle in free space.
The unperturbed states are plane waves, characterized by
a momentum ${\bf \vec{k}}$,
and an energy $\epsilon_{\bf \vec{k}} = | {\bf \vec{k}}|^2 / M$.
The inverse lifetime of this state, when the 
coupling to the environment is described by the action in eq.(\ref{action}),
is:
\begin{widetext}
\begin{equation}
\Gamma_{\bf \vec{k}} ( T )  = \int d \omega \int d {\bf \vec{q}}
{\cal F} ( {\bf \vec{q}} ) | \omega | \left[
\left( 1 + e^{- \omega / T} \right) \delta \left( 
\omega - \epsilon_{\bf \vec{k}} + \epsilon_{\bf \vec{k} + \vec{q}}
\right) +  e^{- \omega / T} \delta \left(
\omega + \epsilon_{\bf \vec{k}} - \epsilon_{\bf \vec{k} + \vec{q}}
\right) \right]
\label{lifetime}
\end{equation}
\end{widetext}
where ${\cal F} ( {\bf \vec{q}} )$ is the Fourier transform
of ${\cal F} \left[
{\bf \vec{X}} ( \tau ) - {\bf \vec{X}} ( \tau' ) \right]$
in eq.(\ref{action}). The two terms in the r.h.s. of eq.(\ref{lifetime})
describe the emission and absorption of quanta by the
environment.

For the case of a dirty electron gas studied earlier,
${\cal F} ( {\bf \vec{q}} ) \approx 1 / ( | {\bf \vec{q}} |^2 \nu {\cal D} )$,
where $\nu$ is the density of states, and ${\cal D}$ is the diffusion
coefficient\cite{G02}. The integral in eq.(\ref{lifetime})
is convergent, except in one dimension. We can define
selfconsistently a dephasing lifetime\cite{AAK82,SAI90,AAG99}
where contributions from low energy transitions are suppressed. 
Then, when $\epsilon_{\bf \vec{k}} \ll T$, we find:
\begin{equation}
\lim_{T \rightarrow 0}
\Gamma_{\bf \vec{k}} ( T ) \approx \frac{\hbar}{M \nu {\cal D} l_T^D}
\propto T^{D/2}
\label{lifetime_T}
\end{equation}
where $l_T = \sqrt{\hbar^2 / ( M T )}$. There are logarithmic corrections
for $D=1$, which arise from the divergence
in the integral in eq.(\ref{lifetime}).

We can also calculate the real part of the self energy, which,
at zero temperature, is:
\begin{equation}
{\rm Re} \Sigma \left( {\bf \vec{k}} , \omega \right)
= \int d \omega' \int d {\bf \vec{q}}
\frac{{\cal F} ( {\bf \vec{q}} )
| \omega' |}{\omega - \omega' + \epsilon_{\bf \vec{k}}
- \epsilon_{{\bf \vec{k}} + {\bf \vec{q}}}}
\end{equation}
Quantities associated to the derivatives
of the real part of the self energy, like  
the effective mass renormalization and the quasiparticle residue:
\begin{eqnarray}
\frac{\delta M}{M^2} &= &- \left. \nabla^2_{\bf \vec{k}}
{\rm Re} \Sigma \left( {\bf \vec{k}} , \omega 
\right) \right|_{\omega = \epsilon_{\bf \vec{k}}} \nonumber \\
Z &= &\left[ 1 + \left. \frac{\partial {\rm Re}
\Sigma \left( {\bf \vec{k}} , \omega  \right)
}
{\partial \omega} \right|_{\omega = \epsilon_{\bf \vec{k}}}
 \right]^{-1}
\end{eqnarray}
are divergent as ${\bf \vec{k}} \rightarrow 0$, for D=1,2.

 Eq.(\ref{lifetime_T}) suggests that quantum effects are still
important at zero temperature. This interpretation, however, is
not consistent with the non perturbative results for $D=1,2$
(see subesections IVD and IVE). In both cases, we find that $G ( \tau )$
increases more slowly than $\tau$, so that the energy scale
for the Aharonov-Bohm oscillations in a closed orbit
acquire an anomalous $R$ dependence (see IVB).
We can infer an effective ^^ ^^ dephasing time", from the scale
at which $G ( \tau )$ reaches its asymptotic unconventional
behavior. In both cases, this scale is $\tau_\phi
\propto ( \hbar^2 / M l^2 )^{-1}$,
where $l$ is the mean free path in the environment.
The unperturbed case is recovered when $l \rightarrow \infty$.

The Caldeira-Leggett model leads to ${\cal F} ( {\bf \vec{q}} )
\approx \delta'' ( {\bf \vec{q}} )$, 
and, 
for $\epsilon_{\bf \vec{k}} \ll T$,
$\lim_{T \rightarrow 0}
\Gamma_{\bf \vec{k}} ( T ) \approx \frac{\hbar \eta}{M}
$. 
This result is consistent with the non perturbative,
analysis of the model, which suggests that quantum effects are
strongly suppressed, even at zero temperature.

\section{Conclusions}
We have analyzed the low energy properties of a quantum particle
interacting with dissipative environments, characterized by an
ohmic response. By means of a large-$N$, or variational approximation,
and numerical calculations, we have estimated time correlations 
which characterize the dynamics of the particle. 
Our method allows us to treat both the weak coupling limit,
and the Caldeira-Leggett model, 
where the coupling to the environment strongly suppresses
the quantum properties of the particle.
The retarded interactions induced by the
coupling to the environment are described by a function,
${\cal F} \left[ | {\bf \vec{X}} ( \tau ) - {\bf \vec{X}} ( \tau' )
|^2 \right]$, which depends on the type of environment, eq.(\ref{action}). 
For a particle propagating freely, 
we characterize the dynamics of the particle
by the correlation function $G (  \tau - \tau' ) = \left\langle
\left[ {\bf \vec{X}} ( \tau ) - {\bf \vec{X}} ( \tau' ) 
\right]^2 \right\rangle$. 

We find that ohmic environments can be divided into two broad classes
(even if they all give rise to the same macroscopic dissipative
equations of motion):

i) If $\lim_{u \rightarrow \infty} {\cal F}' ( u )$ decays faster
than $u^{-1}$, the behavior of the particle can be described in terms
of an effective mass, $M_{eff}$. 
At long times, $G ( \tau ) \propto
M_{eff} \tau$. We have estimated the renormalization of the
bare mass, which depends exponentially on the coupling.
Environments with these features are the clean and dirty
three dimensional electron liquids, and the short range kernel where
${\cal F} ( u ) \propto 1 - e^{-u / l}$.

ii) If $\lim_{u \rightarrow \infty} {\cal F}' ( u )$ decays more
slowly that $u^{-1}$, the effective mass of the particle becomes
infinite at zero temperature. At long times,
$G ( \tau ) \propto \log^\gamma ( \tau )$. This is the behavior of 
the Caldeira-Leggett model ($\gamma=1$), and the one dimensional dirty
electron gas ($\gamma=1/2$).

The two dimensional dirty electron gas is an intermediate case.
Our results are consistent with a power law dependence,
$G ( \tau ) \propto
\tau^\kappa$, with $0 < \kappa < 1$, and $\kappa \propto 1 - g^{-1}$,
where $g$ is the conductance.

We have also considered the properties of a particle in an external potential,
or moving around a closed ring. 
We find different qualitative behavior corresponding to the same
two cases discused above:

i) If $\lim_{u \rightarrow \infty} {\cal F}' ( u )$ decays faster
than $u^{-1}$, a localized potential must exceed a threshold strength
before localized solutions are possible. The 
energy scale which characterize the quantum properties of the particle
is $\hbar^2 / ( M_{eff} R^2 )$, for $R \gg l$, where $l$ is a length
scale which describes the range of the fluctuations in the environment.
A model characterized by a function ${\cal F}$ which fluctuates
between zero and a finite value
as $u \rightarrow \infty$ can be included in this class\cite{GS85}.

ii) If $\lim_{u \rightarrow \infty} {\cal F}' ( u )$ decays more
slowly that $u^{-1}$, a localized solution exists for arbitrarily
weak confining potentials. The characteristic
energy scale which defines the quantum properties around a
ring decays, for $R \gg l$, as $e^{- c (R/l)^{2 \gamma}}$, where $c$ 
and $\gamma$ are constants.

In some cases, there is a first order phase transition, as function
of the strength of the potential, or the size of the orbit, between
a weakly and a strongly localized solution.

Summarizing, we find that different ohmic environments can influence the 
quantum properties of an external particle in different ways:
couplings with long range spatial interactions, such as
in the Caldeira-Leggett model, strongly suppress quantum effects,
while the effects of less singular couplings can be qualitatively
understood within perturbation theory.
\section{Acknowledgements}
I am thankful to D. Arovas, M. B\"uttiker, C. Herrero,
R. Jalabert, A. Kamenev, F. Sols and A. Zaikin
for helpful comments and suggestions. 
I acknowledge financial support from grants
PB96-0875 (MCyT, Spain), and 07N/0045/98 (C. Madrid).

\end{document}